\documentclass[pre,twocolumn,superscriptaddress,tightenlines,showpacs,longbibliography,floatfix,footinbib,preprintnumbers]{revtex4-1}

\newcommand{\preprintid}{RBI-ThPhys-2026-11}
\preprint{\preprintid}

\usepackage[T1]{fontenc}
\usepackage{latexsym,amsmath,amsfonts,tabularx,amssymb,bm,empheq}

\usepackage{subfigure}

\usepackage{hyperref}
\usepackage{blkarray}
\usepackage{mathtools}
\usepackage{multirow}
\usepackage{microtype}
\usepackage{tensor}
\usepackage{tensor}
\usepackage{bbm}
\usepackage{braket}

\usepackage{color}
\usepackage[dvipsnames]{xcolor}
\definecolor{cgreen}{RGB}{76,153,0}
\definecolor{myorange}{RGB}{245,156,74}

\usepackage{fancyhdr}
\fancypagestyle{preprintfirstpage}{%
  \fancyhf{}%
  \fancyfoot[L]{\preprintid}%
}

\begin{document}

\title{Cell divisions suppress dynamical correlations in solid tissues}

\author{Ali Tahaei}
\affiliation{Max Planck Institute for the Physics of Complex Systems, 	N\"{o}thnitzer Str.38, 01187 Dresden, Germany}

\author{Ahandeep Manna}
\affiliation{Max Planck Institute for the Physics of Complex Systems, 	N\"{o}thnitzer Str.38, 01187 Dresden, Germany}

\author{Marko Popović}
\affiliation{Max Planck Institute for the Physics of Complex Systems, 	N\"{o}thnitzer Str.38, 01187 Dresden, Germany}
\affiliation{Center for Systems Biology Dresden, Pfotenhauerstrasse 108, 01307 Dresden, Germany}
\affiliation{Theoretical Physics Division, Ru\dj er Bošković Institute, Bijenička cesta 54, 10000 Zagreb, Croatia}

\begin{abstract}
Developing tissues often maintain mechanical coherence while continuously remodeling through cellular processes such as cell divisions and rearrangements. In this way, they are an example of amorphous solids. In passive amorphous solids, local rearrangements can trigger one another through long-ranged elastic interactions, leading to system-spanning avalanches near yielding. Whether similar collective dynamics should be expected in living tissues is unclear, because cell divisions generate stress and remodeling events independently of local mechanical stability. Here, we address this question using a two-dimensional elastoplastic model in which cell divisions are treated as active plastic events. We find that while cell divisions fluidize the tissue below the passive yield stress, but preserve the marginal stability in the quasistatic limit. However, they also strongly suppress the system-spanning avalanches of cell rearrangements, in constrast with the expected behavior in passive amorphous solids. Finally, we show that the avalanche supression originates from the energy balance in the system. Namely, the energy injected by cell divisions allows for shear flow below the yield stress, but also provides a finite budget for rearrangements. These results suggest that proliferating tissues display the structural hallmarks of marginal amorphous solids while exhibiting much shorter-ranged correlations in dynamics, compared to passive amorphous solids.
\end{abstract}

\maketitle
\thispagestyle{preprintfirstpage}

\section{Introduction}

The emergence of shape in developing tissues requires materials that are mechanically coherent enough to transmit forces, yet able to remodel over time. This remodeling occurs through localized cellular processes such as neighbor exchanges, extrusions, and divisions \cite{Heisenberg2013,LeGoff2016,Julicher2017}. Neighbor exchanges and extrusions have close analogs in passive disordered materials, where particles can rearrange or break. Cell divisions are more distinctive of living systems: they create active remodeling events with no direct passive counterpart, while still deforming the surrounding material and contributing to tissue-scale flows \cite{Ranft2010,Matoz-Fernandez2017}. This raises the broader question of how such active remodeling events reshape the mechanics of a disordered solid.

In passive amorphous solids, localized plastic events interact through long-ranged elastic fields. A rearrangement in one region redistributes stress throughout the material and can trigger further rearrangements, leading to system-spanning avalanches near yielding \cite{Picard2004,Nicolas2018,Lin2014a,Mueller2015,Lin2015,Lin2016a}. This collective dynamics can be shown to arise from  marginal stability of the system in presence of long-ranged interactions \cite{Mueller2015}. In a living material, however, cell divisions add effective plastic events independent of the local stability. It is therefore unclear whether such activity preserves the avalanche phenomenology and even the marginal stability of amorphous solids.

Here, we address this question with a two-dimensional elastoplastic model, appropriate for solid-like epithelial tissues. Recent analysis of experimental data showed that a cell division, at a coarse-grained level, imprints an anisotropic force dipole in the surrounding elastic tissue \cite{Tahaei2025}. We therefore treat divisions as active plastic events and compare randomly oriented divisions with divisions that relax the local stress or align with it. We find that divisions fluidize the tissue below the passive yield stress, consistent with previous findings \cite{Ranft2010,Matoz-Fernandez2017}, while maintaining the marginal stability in the quasistatic limit. Interestingly, we observe that despite marginal stability and long-range elastic interactions, avalanches of cell rearrangements become strongly supressed. We show that this is a consequence of the energy balance in the system: the energy injected by cell divisions allows for shear flow below the yield stress, but also provides a finite budget for rearrangements. These results suggest that proliferation can preserve the local mechanical organization of a marginal amorphous solid while actively limiting the range of collective rearrangements. Our work provides a possible explanation for why epithelial tissues can show glassy mechanical signatures without exhibiting the system-spanning avalanches expected in a passive amorphous solid.

\section{The elasto-plastic model with cell divisions}
The elastoplastic model (EPM) is a standard coarse-grained description of yielding in amorphous solids, in which the material responds elastically between localized plastic events \cite{Nicolas2018,Baret2002,Jagla2007,Jagla2020}. Examples of such events include shear transformation zones in glasses \cite{Argon1979,Lemaitre2009} and T1 transitions in dry foams and epithelial tissues \cite{Farhadifar2007,Fletcher2014}. Here, we use this framework to study cell divisions as active plastic events, extending earlier mean-field work that considered only randomly oriented divisions \cite{Matoz-Fernandez2017}.

\begin{figure*}
    \centering
    \includegraphics[width=.75\linewidth]{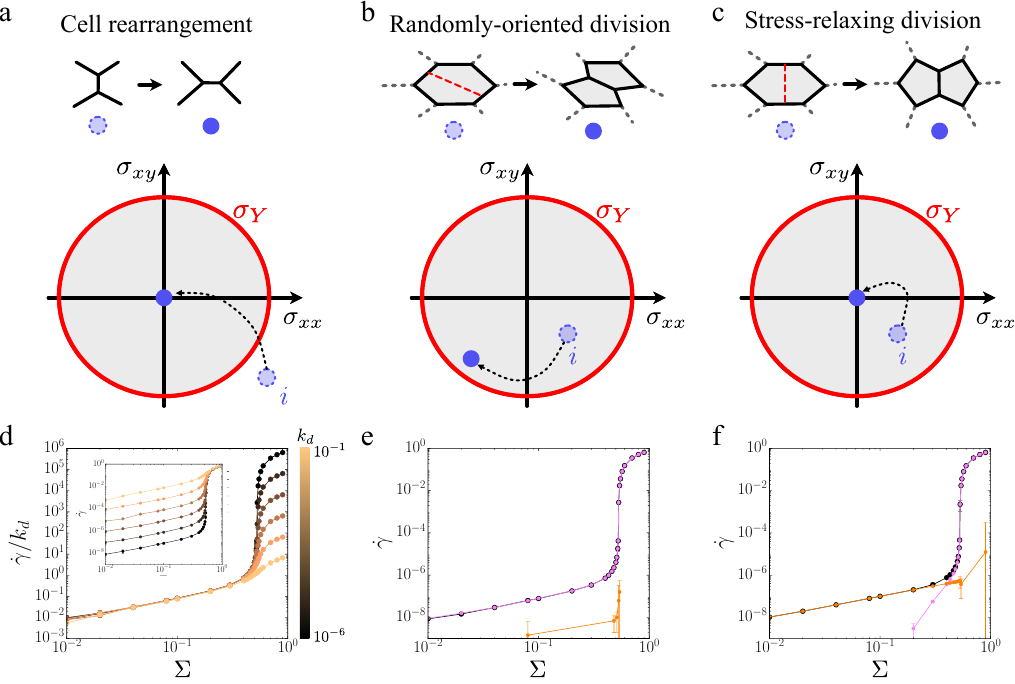}
    \caption{Cell divisions fluidize an elastoplastic sheet below its passive yield stress. (a-c) Schematic of local events in the EPM. (a) When a block becomes mechanically unstable, ${\sigma}^2_i \geq \sigma_Y^2$, it undergoes a plastic event and relaxes its local shear stress, ${\bm{\sigma}}_i \to 0$. (b) Randomly oriented divisions imprint a bounded shear increment $\bm{\Delta}_{\mathrm{div}}$ whose orientation is independent of the local stress. (c) Stress-relaxing divisions reset the local stress in the same symmetry channel as a passive plastic event. In both cases, divisions are controlled by the cell cycle rather than by mechanical instability. (d) Flow curves showing that divisions generate a linear response below $\Sigma_c$, while the passive rheology is recovered above $\Sigma_c$. The inset shows the collapse obtained by rescaling the shear rate by the division rate $k_d$. The plot shows the simulation results for randomly-oriented cell divisions. (e,f) Decomposition of the total shear rate into a division contribution $C$ and a rearrangement contribution $R$. For randomly oriented divisions, the flow is carried almost entirely by rearrangements, whereas for stress-oriented divisions, the low-stress flow is dominated by the divisions themselves. In all EPM simulations, the imposed stress is $\Sigma_{xy}$, the division rate is $k_d=10^{-6}$, and the system size is $L=128$.}
    \label{fig:DivisionRate-Flow}
\end{figure*}

In the elastoplastic model, the system is divided into $N$ mesoscopic blocks, and the state of each block is described by its local shear stress $\bm{ \sigma}{}_i$. In two-dimensional model, the shear stress tensor  has two independent components $\sigma_{xx,i}$ and $\sigma_{xy,i}$. To describe cell rearrangements, a yield stress surface is defined in each block, which we choose to be  a circle, as shown in Fig.~\ref{fig:DivisionRate-Flow}. Radius of the circle we denote as the local yield stress $\sigma_Y$ and keep constant in all blocks.
If the shear stress in a block is beyond the yield surface, the block becomes unstable, which is modelled by a probability rate $1/\tau_0$ for the block to experience a cell rearrangement and relax. Here, $\tau_0$ represents a microscopic relaxation time-scale. Upon relaxation, the shear stress is reduced to a low magnitude value and the corresponding plastic deformation in the block is accumulated $\Delta {\bm{\epsilon}}_p= { \bm{\sigma}}_i/\kappa$, where $\kappa$ is the local shear modulus, see Methods for further details.

Following a cell rearrangement in a block, shear stress in the system has to be redistributed to maintain the force balance as
\begin{align}
    \bm {\sigma}{}_j \to \bm {\sigma}{}_j  + \kappa \mathbf{G}(\Delta \bm{r}_{ij}) \Delta \bm {\epsilon}{}_i^{pl},
    \label{eq:stress_redist_plEvent}
\end{align}
where $\mathbf{G}$ is the linear elastic propagator of elastic stress induced by a force dipole in two-dimensional sheet \cite{Eshelby1957,Nicolas2018}, $\Delta \bm{r}_{ij}=\bm{r}_i - \bm{r}_j$, and $\kappa$ is the shear modulus.
In Fourier space, the elements of the propagator are given by,
\begin{align}
    \widehat {G}_{xx,xx} &= - \frac{(q_x^2-q_y^2)^2}{(q_x^2+q_y^2)^2} \\
    \widehat {G}_{xy,xy} &= - \frac{4q_x^2q_y^2}{(q_x^2+q_y^2)^2} \\
    \widehat {G}_{xx,xy} = \widehat {G}_{xy,xx} &= - \frac{2q_xq_y(q_x^2-q_y^2)}{(q_x^2+q_y^2)^2}
\end{align}
with $q_x,q_y$ the Fourier modes.
In a discrete system with bi-periodic boundary conditions, there is a correction to the Fourier modes described as $q_x^2=2(1-\cos(2\pi n_x/L))$, $q_y^2=2(1-\cos(2\pi n_y/L))$ and $q_xq_y=2\sin(2\pi n_x/L) \sin(2\pi n_y /L)$ where $n_\alpha = -L/2+\lbrace 1, \cdots, L \rbrace$ with $n=x,y$.

We extend the EPM to describe epithelial tissues by treating cell divisions as active plastic events whose remodeling remains imprinted in the tissue as an anisotropic force dipole. We have recently shown that this description reproduces experimentally measured strain fields in the developing \textit{Drosophila} wing epithelium and that the division-induced dipole has the same symmetry as the dipole generated by a cell rearrangement~\cite{Tahaei2025}. In the model, divisions occur independently in all blocks with rate $k_d$. Each division accumulates a plastic strain $\Delta {\bm{\epsilon}}_i^d$ and changes the local shear stress as $\bm {\sigma} {}_i \to \bm {\sigma} {}_i - \kappa \Delta \bm{\epsilon}{}_i^d$. We distinguish three classes of divisions. Randomly oriented divisions (ROD) have no preferred axis and draw their strain magnitude from a bounded distribution. Stress-oriented divisions (SOD) have the same magnitude statistics but align their axis with the local shear stress. Stress-relaxing divisions (SRD) relax the local stress in the same symmetry channel as a passive plastic event. The two former classes are motivated by Hertwig's law in proliferating epithelia~\cite{Hertwig1884, Bosveld2016, Tahaei2025}. In the main text, we focus on ROD and SRD and report SOD results in the Appendix. Further implementation details are given in Appendix~\ref{appendix:division_implementation}.

\section{Tissue fluidisation by cell divisions}
In the absence of activity, the steady-state rheology of the EPM is described by a Herschel-Bulkley relation, $\dot{\gamma}=A[\Sigma-\Sigma_c]^\beta$, where $\dot{\gamma}$ is the shear rate component along the imposed stress $\Sigma$ and $\Sigma_c$ is the yield stress \cite{Herschel1926,Nicolas2018}. Cell divisions introduce an additional timescale, $1/k_d$, and therefore provide a natural mechanism for flow below the passive yield stress. We simulate the steady state flow of the EPM with cell divisions and, Consistent with earlier work \cite{Ranft2010,Matoz-Fernandez2017}, we find a linear response $\dot{\gamma}\sim \Sigma$ at low stress, with a viscosity proportional to $1/k_d$; see Fig.~\ref{fig:DivisionRate-Flow}. At stresses above $\Sigma_c$, the division rate becomes negligible compared to the rate of mechanically triggered rearrangements, and the passive non-linear rheology is recovered.

To further characterize the flow, we decompose the plastic strain rate into a contribution from cell divisions, $C$, and a contribution from cell rearrangements, $R$. For randomly oriented divisions, the direct shear produced by cell divisions averages to zero, so the macroscopic flow is almost entirely carried by cell rearrangements. By contrast, for stress-oriented or stress-relaxing divisions, the low-stress flow is dominated by the divisions themselves, while the cell rearrangement contribution becomes negligible, see Fig.~\ref{fig:DivisionRate-Flow}. Therefore, although the overall linear flow below $\Sigma_c$ is present independent of the cell division type, the actual origin of the flow can vary significantly between being entirely due to cell rearrangements to being entirely due to cell divisions as the type of the cell division is changed. To better understand how the flow structure depends on the division type, we next turn to characterizing the stability of the tissue in the quasistatic limit.

\section{Marginal stability}
We consider the quasistatic cell division rate limit $k_d\to 0^+$, where the system is fully relaxed after each division before the next division occurs. In this limit, the relevant observable in each EPM block is the local stability $x$, corresponding to the additional stress required to reach the local yield surface. In passive amorphous solids, the distribution of local stability $P(x)$ has been shown to characterize the  state and rheology of amorphous solids that develop a pseudogap, $P(x)\sim x^\theta$, with $\theta = 0.55\pm 0.03$ for randomly oriented divisions and $\theta= 0.52\pm 0.04$ for stress-relaxing divisions, see Appendix B, both consistent with values near $\theta\approx 0.57$ reported in literature \cite{Karmakar2010, Lin2014b, Lin2015, Lin2016a}. Such a pseudogap distribution is characteristic of marginally stable systems with long-ranged elastic interactions \cite{Lin2014a,Mueller2015}.

\begin{figure}
    \centering
    \includegraphics[width=\linewidth]{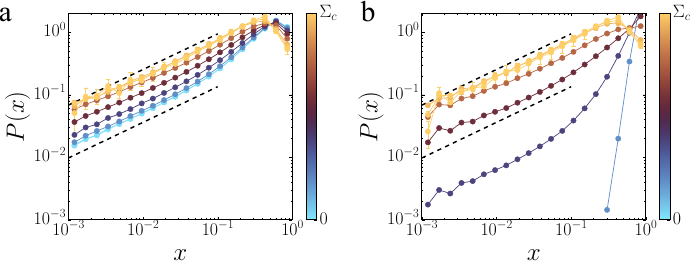}
    \caption{Distribution of local stability $P(x)$ at different imposed shear stresses ${\Sigma}$ for two division types: (a) randomly oriented and (b) stress-relaxing. In both cases, the data are consistent with a pseudogap $P(x)\sim x^\theta$, with $\theta=0.55\pm0.03$ for randomly oriented divisions and $\theta=0.52\pm0.04$ for stress-relaxing divisions. For stress-relaxing divisions, the low-$x$ tail is progressively depleted as the distribution narrows and the tissue becomes more stable. All simulations use system size $L=128$ and $\Delta_{\rm{di}}=1$.}
    \label{fig:EPM_division_pseudogap}
\end{figure}

For randomly oriented divisions, we find that the tissue remains marginally stable throughout the flowing phase below $\Sigma_c$. The distribution $P(x)$ retains a robust pseudogap with exponent $\theta$, independent of stress and consistent with two-dimensional amorphous solids at $\Sigma_c$, see Fig.~\ref{fig:EPM_division_pseudogap}a. Stress-relaxing divisions lead to a different phenomenology. As the imposed stress is reduced, the distribution first still appears compatible with a pseudogap, but dramatically depleted as the imposed stress is reduced, see Fig.~\ref{fig:EPM_division_pseudogap}b. Then, for the lowest stress, we consider that we do not even observe $x$ in the regime where the power-law scaling is expected. This raises the question of whether the system remains marginally stable or whether a true gap opens in $P(x)$.

\begin{figure}
    \centering
    \includegraphics[width=.9\linewidth]{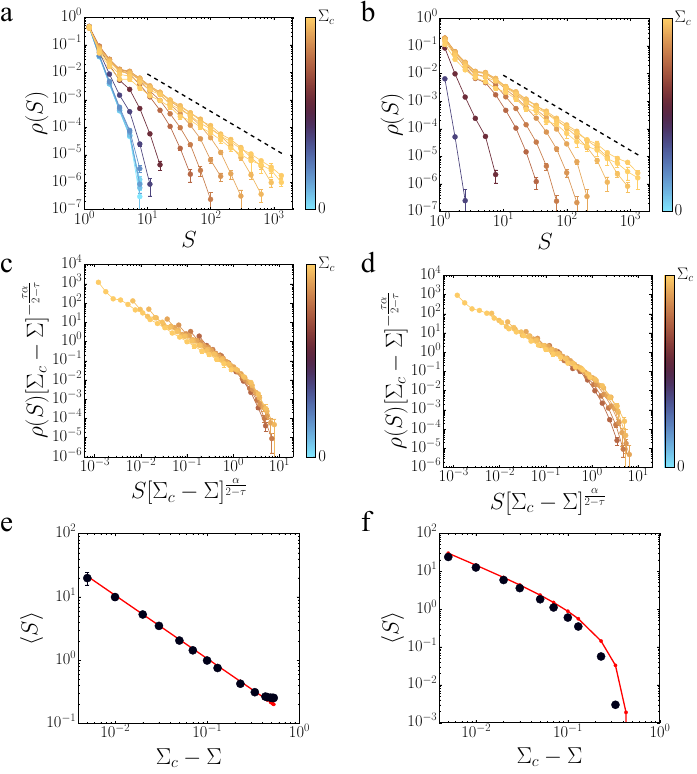}
    \caption{The distribution of avalanche size $\rho(S)$ of different types of cell divisions, (A) randomly-oriented divisions, (B) Stress-relaxing divisions, at various imposed shear stress ${\Sigma} \in [0, \Sigma_c]$. The left panel on each row (\textit{i}) compares the distribution to a power law decay $\rho(S) \sim S^{-\tau}$ with $\tau=1.36$ shown by the black dashed line. (\textit{ii}) A successful collapse of curves with ${\Sigma}>0.8\Sigma_c$ is achieved by considering a cut-off avalanche size $S_c\sim[\Sigma_c-{\Sigma}]^{\alpha/(2-\tau)}$ where we find $\alpha=1.00\pm0.05$. In all simulations the system size is $L=128$ and $\Delta_{\rm{div}}=1$. (e,f) The average avalanche size $\langle S \rangle$ against $\Sigma_c -{\Sigma}$ for randomly-oriented divisions (e) and stress-relaxing divisions (f). The red line shows the predicted $\langle S \rangle$ obtained from the elastic energy balance between cell divisions and rearrangements, see Eqs.~\ref{eq:general_elasitc_E_div}, \ref{eq:deltaE_re}, \ref{eq:deltaE_ROD}, and \ref{eq:deltaE_SRD}. In all simulations, the box size is $L=256$ and $\Delta_{\rm{div}}=1$.}
    \label{fig:EPM_division_avalanches}
\end{figure}

To understand the depletion of $P(x)$, we analyze the mean-field H\'ebraud-Lequeux model  (HL) \cite{Hebraud1998,Agoritsas2015} with cell divisions \cite{Matoz-Fernandez2017} in the limit $\Sigma=0$, see Appendix \ref{appendix:Hebraud-Lequeux}. 
For randomly oriented divisions, we find the usual linear pseudogap of the H\'ebraud-Lequeux class, $P(x)\sim x/D_m$, where $D_m$ is the mechanical noise parameter. For stress-relaxing divisions, the distribution still vanishes linearly, but its prefactor is reduced as a function of $D_m$, with $P(x)\sim A x$ where $A\sim e^{-1/\sqrt{D_m}}/D_m$. 
In other words, the system remains formally marginal, yet the number of blocks that are actually available to yield becomes is reduced. Estimating the typical lowest $x$ in the system of size $N$ from $1/N=\int_0^{x_{\min}}P(x)\,dx$ gives $x_{\min}^{\rm ROD}\sim \sqrt{D_m/N}$ for randomly oriented divisions, whereas $x_{\min}^{\rm SRD}\sim \sqrt{D_m e^{1/\sqrt{D_m}}/N}$ for stress-relaxing divisions. This corresponds to a significant shift of the weakest block away from instability at $x= 0$, with $e^{1/\sqrt{D_m}}$ estimated to be of the order $10-30$, see Appendix \ref{appendix:Hebraud-Lequeux} for the estimate.

This picture also provides an intuitive interpretation. As the value of stress in a block diffuses, without drift at $\Sigma= 0$, to eventually reach the yield surface, stress-relaxing divisions repeatedly reset local stresses. In the HL model, this is effectively a diffusion-evaporation dynamics, which gives rise to the exponentially decaying density of states. In practice, stresses rarely manage to diffuse to the yield surface before another division occurs, resetting the stress to $0$. The tissue therefore formally retains marginal stability, but the pool of plastically available regions is strongly depleted, and in a finite system, it is possible that the typical smallest $x_\text{min}$ be higher than the regime where the pseudogap scaling is expected. 
Although in the two-dimensional system the mechanical noise is more complex, leading to stress dynamics different from a simple diffusion, the evaporation mechanism by stress-relaxing divisions is the same, i.e., the probability of survival exponentially decreases with time, but is never exactly $0$. 
Therefore, although our numerical measurements could not reach the scaling regime for the lowest stress value, we do not expect to find a true gap, as in a big enough system, there will always be some blocks where no division occurred before reaching the yield surface. 

\section{Cell divisions suppress avalanches}
In passive amorphous solids, marginal stability and long-ranged elastic interactions are accompanied by system-spanning avalanches of particle rearrangements \cite{Lin2014a,Mueller2015,Lin2015,Lin2016a}. To test whether this remains true in a proliferating tissue, we consider the steady state flow in the quasistatic cell division rate limit and define the avalanche size $S$ as the number of cell rearrangements triggered after each division. The corresponding distributions $\rho(S)$ are shown in Fig.~\ref{fig:EPM_division_avalanches}. At $\Sigma=\Sigma_c$ they are consistent with a power-law form $\rho(S)\sim S^{-\tau}$ with $\tau=1.36\pm0.01$ for both division types, in agreement with two-dimensional elastoplastic models. Below $\Sigma_c$, however, the distributions acquire a stress-dependent cutoff $S_c$ that decreases as the imposed stress is reduced. For stress-relaxing divisions, this cutoff becomes so severe at low stress that avalanches are effectively absent. To quantify the cutoff, we use the scaling form
\begin{align}
    \rho(S) \sim S^{-\tau} f\left( \frac{S}{S_c} \right), \label{eq:division_S_scaling_thesis}
\end{align}
with $S_c \sim [\Sigma_c - \Sigma]^{\alpha/(2-\tau)}$, where $\alpha$ is a free parameter. This scaling function is able to collapse the avalanche distributions with $\alpha=1.00 \pm 0.05$ independent of the division type.

To understand the origin of the cutoff, we consider the elastic energy balance $E=\sum_i {\sigma}_i^2/2\kappa$ and set $\kappa=1$ without a loss of generality. In the steady state, the energy injected by cell divisions and external loading must compensate for the energy dissipated through cell rearrangements
\begin{align}
    \langle S \rangle \mathbf{E}[\delta E^{\rm{re}}] + \mathbf{E}[\delta E^{\rm{div}}] = 0
    \label{eq:general_elasitc_E_div}
\end{align} 
where $\mathbf{E}[\cdots]$ denotes the average over an ensemble of events, and we accounted for the fact that for each cell division there are on average $\langle S \rangle$ cell rearrangements.
As the stress $\Sigma$ approaches $\Sigma_c$ from below, the amount of energy disspated by the cell rearrangements vanishes linearly with the stress distance, 
\begin{align}
    \mathbf{E}[\delta E^{\rm{re}}] = -(\Sigma_c - {\Sigma}) \mathbf{E}[{\sigma}_m], \label{eq:deltaE_re}
\end{align}
where $\mathbf{E}[{\sigma}_m]$ is the average shear stress of blocks at the moment when they experience a cell rearrangement, which is bounded from below by $\sigma_Y$, see Appendix \ref{appendix:elasticEnergy}. By contrast, the average injection of energy through cell divisions $\mathbf{E}[\delta E^{\rm{div}}]$ remains finite at $\Sigma_c$. More specifically, for randomly-oriented and stress-relaxing divisions, we find
\begin{align}
    \mathbf{E}[\delta E^{\rm{div}}_{\rm{ROD}}] &= \left(\frac{1}{2} + D\right)\eta_{\text{div}}^2, \label{eq:deltaE_ROD}\\
    \mathbf{E}[\delta E^{\rm{div}}_{\rm{SRD}}] &= -(\frac{1}{2}-D)\mathbf{E}[{\sigma}^2]+{\Sigma}^2, \label{eq:deltaE_SRD}
\end{align}
respectively. $D$ represents the magnitude of the mechanical noise, which is set by the linear elastic propagator $\mathbf{G}$, and $\eta_{\text{div}}^2$ is the variance of stress generated by cell divisions, see Appendix \ref{appendix:division_el_balance}. 
This allows us to estimate the average avalanche size $\langle S \rangle$ from the energy balance to find $\langle S \rangle \sim (\Sigma_c - \Sigma)^{-1}$ in both cases. Thus, the avalanche size cutoff scales 
\begin{align}
    S_c \sim [\Sigma_c-\Sigma]^{1/(2 - \tau)}
    \label{eq:cut_off-division}
\end{align}
where we used $\langle S \rangle \sim S_c^{2-\tau}$. This scaling is consistent with our two-dimensional simulations. For randomly oriented divisions, the scaling is independent of ${\Sigma}$, see Fig.~\ref{fig:EPM_division_avalanches} (e). For stress-relaxing divisions, the same mechanism is more extreme, since blocks with low $x$ are already depleted, even triggering any cell rearrangement by a cell division becomes rare, and $\langle S \rangle$ quickly drops to zero as shown in Fig.~\ref{fig:EPM_division_avalanches} (f).

\section{Discussion}
We studied how cell divisions reshape the mechanics of a solid-like tissue when they are described as active plastic events coupled by long-ranged elasticity, following our recent work \cite{Tahaei2025}. This work identifies three main properties of the amorphous solid flowing in the presence of cell divisions. First, divisions fluidize the material below the yield stress of the passive system and generate a linear low-stress response controlled by the division rate. This fact was already established in previous work \cite{Ranft2010,Matoz-Fernandez2017}, but we find that the type of the flow structure depends on the division mechanism. Second, we find that mechanical noise induced by the cell divisions does not open a true gap, unlike the thermal noise in a passive system \cite{Popovic2021Thermal}. Therefore, in the limit of vanishing cell division rate, the system remains marginally stable, although stress-relaxing divisions may significantly suppress the distribution $P(x)$, i.e., the amount of available plasticity. Third, we have shown that, unlike in passive system \cite{Mueller2015}, in the presence of cell divisions, a marginally stable system with long-range interactions does not need to exhibit system-spanning avalanches. Instead, they are limited in size by the energy injected through the cell divisions.

%Our results sharpen this distinction by showing that marginal stability and system-spanning avalanches can be decoupled in active matter. For randomly oriented divisions, similar as a passive elastoplastic solid, the density of weak regions retains a pseudogap with the exponent value that is stress-independent within the precision of our numerical measurements. For stress-relaxing divisions, by contrast, the distribution of weak spots is strongly depleted at low stress. The H\'ebraud-Lequeux analysis suggests that this depletion does not open a true gap. Instead, the system remains marginal, but with a small prefactor corresponding to the very small density of available weak regions.

In passive athermal systems, the solid phase can display critical avalanches during transient loading \cite{Lin2015}, but it cannot sustain a steady-state flow below $\Sigma_c$. Thermal fluctuations round this transition and allow thermally activated steady flow below yield \cite{Popovic2021Thermal}. Although cell divisions play a similar role in that they also permit a steady flow below $\Sigma_c$, the mechanism is very different. Cell divisions inject elastic energy through discrete active plastic events independent of whether the system is locally weak or not. Thus, unlike thermal activation, which is exponentially more likely to activate  regions of the material close to becoming unstable, division-driven fluidisation is much stronger in magnitude; it preserves the marginal stability and limits the avalanches by the energy budget they supply.

%This distinction helps explain why avalanche phenomenology is so strongly altered. In passive amorphous solids, a pseudogap together with long-ranged elastic interactions implies broad, system-spanning cascades. In the active tissue, however, each division also resets part of the local stress landscape and thereby competes with the accumulation of weak regions. The energy-balance argument then clarifies why the average avalanche size is set by the stress distance to yielding rather than by the system size. Activity therefore cuts off correlations not by screening elastic interactions, but by regulating how much passive plasticity can accumulate between active events.

Therefore, activity can preserve marginal stability while sharply reducing dynamical correlations and system-spanning avalanches. In proliferating tissues, divisions do not simply add noise on top of a passive yielding material.
They continuously reshape the population of regions close to local instability and thereby limit how much passive plasticity in the form of cell rearrangements remains available to organize into collective bursts of plasticity. These results suggest a useful viewpoint for tissue mechanics: a proliferating tissue will appear similar to an amorphous solid at yield stress in terms of the density of local stability $P(x)$, consistent with the observation in the fruit fly wing \cite{Popovic2021a}, yet exhibit a significantly less correlated dynamics compared to a passive counterpart.
This may help us understand why epithelial tissues have been reported to show signatures of glassy rheology \cite{Harris2012, Schotz2013}, while normally lacking system-spanning avalanches. 

It will be interesting to include other sources of active noise, such as mechanical fluctuations induced by the stochastic dynamics of molecular motors, which would appear more similar to the thermal noise \cite{Curran2017,Bi2016,Yamamoto2022}. In such a system, it is at the moment unclear whether cell divisions or thermal-like noise would determine the stability and dynamics of plasticity.

\acknowledgements
AM acknowledges the Max Planck school Matter to Life.

\bibliography{bib.bib}

\begin{thebibliography}{35}%
\makeatletter
\providecommand \@ifxundefined [1]{%
 \@ifx{#1\undefined}
}%
\providecommand \@ifnum [1]{%
 \ifnum #1\expandafter \@firstoftwo
 \else \expandafter \@secondoftwo
 \fi
}%
\providecommand \@ifx [1]{%
 \ifx #1\expandafter \@firstoftwo
 \else \expandafter \@secondoftwo
 \fi
}%
\providecommand \natexlab [1]{#1}%
\providecommand \enquote  [1]{``#1''}%
\providecommand \bibnamefont  [1]{#1}%
\providecommand \bibfnamefont [1]{#1}%
\providecommand \citenamefont [1]{#1}%
\providecommand \href@noop [0]{\@secondoftwo}%
\providecommand \href [0]{\begingroup \@sanitize@url \@href}%
\providecommand \@href[1]{\@@startlink{#1}\@@href}%
\providecommand \@@href[1]{\endgroup#1\@@endlink}%
\providecommand \@sanitize@url [0]{\catcode `\\12\catcode `\$12\catcode `\&12\catcode `\#12\catcode `\^12\catcode `\_12\catcode `\%12\relax}%
\providecommand \@@startlink[1]{}%
\providecommand \@@endlink[0]{}%
\providecommand \url  [0]{\begingroup\@sanitize@url \@url }%
\providecommand \@url [1]{\endgroup\@href {#1}{\urlprefix }}%
\providecommand \urlprefix  [0]{URL }%
\providecommand \Eprint [0]{\href }%
\providecommand \doibase [0]{http://dx.doi.org/}%
\providecommand \selectlanguage [0]{\@gobble}%
\providecommand \bibinfo  [0]{\@secondoftwo}%
\providecommand \bibfield  [0]{\@secondoftwo}%
\providecommand \translation [1]{[#1]}%
\providecommand \BibitemOpen [0]{}%
\providecommand \bibitemStop [0]{}%
\providecommand \bibitemNoStop [0]{.\EOS\space}%
\providecommand \EOS [0]{\spacefactor3000\relax}%
\providecommand \BibitemShut  [1]{\csname bibitem#1\endcsname}%
\let\auto@bib@innerbib\@empty
%</preamble>
\bibitem [{\citenamefont {Heisenberg}\ and\ \citenamefont {Bellaïche}(2013)}]{Heisenberg2013}%
  \BibitemOpen
  \bibfield  {author} {\bibinfo {author} {\bibfnamefont {Carl-Philipp}\ \bibnamefont {Heisenberg}}\ and\ \bibinfo {author} {\bibfnamefont {Yohanns}\ \bibnamefont {Bellaïche}},\ }\bibfield  {title} {\enquote {\bibinfo {title} {Forces in tissue morphogenesis and patterning},}\ }\href {\doibase 10.1016/j.cell.2013.05.008} {\bibfield  {journal} {\bibinfo  {journal} {Cell}\ }\textbf {\bibinfo {volume} {153}},\ \bibinfo {pages} {948–962} (\bibinfo {year} {2013})}\BibitemShut {NoStop}%
\bibitem [{\citenamefont {LeGoff}\ and\ \citenamefont {Lecuit}(2016)}]{LeGoff2016}%
  \BibitemOpen
  \bibfield  {author} {\bibinfo {author} {\bibfnamefont {Loïc}\ \bibnamefont {LeGoff}}\ and\ \bibinfo {author} {\bibfnamefont {Thomas}\ \bibnamefont {Lecuit}},\ }\bibfield  {title} {\enquote {\bibinfo {title} {Mechanical forces and growth in animal tissues},}\ }\href {\doibase 10.1101/cshperspect.a019232} {\bibfield  {journal} {\bibinfo  {journal} {Cold Spring Harbor Perspectives in Biology}\ }\textbf {\bibinfo {volume} {8}},\ \bibinfo {pages} {a019232} (\bibinfo {year} {2016})}\BibitemShut {NoStop}%
\bibitem [{\citenamefont {Jülicher}\ and\ \citenamefont {Eaton}(2017)}]{Julicher2017}%
  \BibitemOpen
  \bibfield  {author} {\bibinfo {author} {\bibfnamefont {Frank}\ \bibnamefont {Jülicher}}\ and\ \bibinfo {author} {\bibfnamefont {Suzanne}\ \bibnamefont {Eaton}},\ }\bibfield  {title} {\enquote {\bibinfo {title} {Emergence of tissue shape changes from collective cell behaviours},}\ }\href {\doibase 10.1016/j.semcdb.2017.04.004} {\bibfield  {journal} {\bibinfo  {journal} {Seminars in Cell \& Developmental Biology}\ }\textbf {\bibinfo {volume} {67}},\ \bibinfo {pages} {103–112} (\bibinfo {year} {2017})}\BibitemShut {NoStop}%
\bibitem [{\citenamefont {Ranft}\ \emph {et~al.}(2010)\citenamefont {Ranft}, \citenamefont {Basan}, \citenamefont {Elgeti}, \citenamefont {Joanny}, \citenamefont {Prost},\ and\ \citenamefont {J{\" u}licher}}]{Ranft2010}%
  \BibitemOpen
  \bibfield  {author} {\bibinfo {author} {\bibfnamefont {J.}~\bibnamefont {Ranft}}, \bibinfo {author} {\bibfnamefont {M.}~\bibnamefont {Basan}}, \bibinfo {author} {\bibfnamefont {J.}~\bibnamefont {Elgeti}}, \bibinfo {author} {\bibfnamefont {J.-F.}\ \bibnamefont {Joanny}}, \bibinfo {author} {\bibfnamefont {J.}~\bibnamefont {Prost}}, \ and\ \bibinfo {author} {\bibfnamefont {F.}~\bibnamefont {J{\" u}licher}},\ }\bibfield  {title} {\enquote {\bibinfo {title} {Fluidization of tissues by cell division and apoptosis},}\ }\href {\doibase 10.1073/pnas.1011086107} {\bibfield  {journal} {\bibinfo  {journal} {Proceedings of the National Academy of Sciences}\ }\textbf {\bibinfo {volume} {107}},\ \bibinfo {pages} {20863–20868} (\bibinfo {year} {2010})}\BibitemShut {NoStop}%
\bibitem [{\citenamefont {Matoz-Fernandez}\ \emph {et~al.}(2017)\citenamefont {Matoz-Fernandez}, \citenamefont {Agoritsas}, \citenamefont {Barrat}, \citenamefont {Bertin},\ and\ \citenamefont {Martens}}]{Matoz-Fernandez2017}%
  \BibitemOpen
  \bibfield  {author} {\bibinfo {author} {\bibfnamefont {D.A.}\ \bibnamefont {Matoz-Fernandez}}, \bibinfo {author} {\bibfnamefont {Elisabeth}\ \bibnamefont {Agoritsas}}, \bibinfo {author} {\bibfnamefont {Jean-Louis}\ \bibnamefont {Barrat}}, \bibinfo {author} {\bibfnamefont {Eric}\ \bibnamefont {Bertin}}, \ and\ \bibinfo {author} {\bibfnamefont {Kirsten}\ \bibnamefont {Martens}},\ }\bibfield  {title} {\enquote {\bibinfo {title} {Nonlinear rheology in a model biological tissue},}\ }\href {\doibase 10.1103/PhysRevLett.118.158105} {\bibfield  {journal} {\bibinfo  {journal} {Physical Review Letters}\ }\textbf {\bibinfo {volume} {118}},\ \bibinfo {pages} {158105} (\bibinfo {year} {2017})}\BibitemShut {NoStop}%
\bibitem [{\citenamefont {Picard}\ \emph {et~al.}(2004)\citenamefont {Picard}, \citenamefont {Ajdari}, \citenamefont {Lequeux},\ and\ \citenamefont {Bocquet}}]{Picard2004}%
  \BibitemOpen
  \bibfield  {author} {\bibinfo {author} {\bibfnamefont {Guillemette}\ \bibnamefont {Picard}}, \bibinfo {author} {\bibfnamefont {Armand}\ \bibnamefont {Ajdari}}, \bibinfo {author} {\bibfnamefont {Francois}\ \bibnamefont {Lequeux}}, \ and\ \bibinfo {author} {\bibfnamefont {Lyderic}\ \bibnamefont {Bocquet}},\ }\bibfield  {title} {\enquote {\bibinfo {title} {Elastic consequences of a single plastic event: a step towards the microscopic modeling of the flow of yield stress fluids},}\ }\href@noop {} {\bibfield  {journal} {\bibinfo  {journal} {The European Physical Journal E}\ }\textbf {\bibinfo {volume} {15}},\ \bibinfo {pages} {371–381} (\bibinfo {year} {2004})}\BibitemShut {NoStop}%
\bibitem [{\citenamefont {Nicolas}\ \emph {et~al.}(2018)\citenamefont {Nicolas}, \citenamefont {Ferrero}, \citenamefont {Martens},\ and\ \citenamefont {Barrat}}]{Nicolas2018}%
  \BibitemOpen
  \bibfield  {author} {\bibinfo {author} {\bibfnamefont {Alexandre}\ \bibnamefont {Nicolas}}, \bibinfo {author} {\bibfnamefont {Ezequiel~E.}\ \bibnamefont {Ferrero}}, \bibinfo {author} {\bibfnamefont {Kirsten}\ \bibnamefont {Martens}}, \ and\ \bibinfo {author} {\bibfnamefont {Jean-Louis}\ \bibnamefont {Barrat}},\ }\bibfield  {title} {\enquote {\bibinfo {title} {Deformation and flow of amorphous solids: Insights from elastoplastic models},}\ }\href {\doibase 10.1103/RevModPhys.90.045006} {\bibfield  {journal} {\bibinfo  {journal} {Reviews of Modern Physics}\ }\textbf {\bibinfo {volume} {90}} (\bibinfo {year} {2018}),\ 10.1103/RevModPhys.90.045006}\BibitemShut {NoStop}%
\bibitem [{\citenamefont {Lin}\ \emph {et~al.}(2014{\natexlab{a}})\citenamefont {Lin}, \citenamefont {Saade}, \citenamefont {Lerner}, \citenamefont {Rosso},\ and\ \citenamefont {Wyart}}]{Lin2014a}%
  \BibitemOpen
  \bibfield  {author} {\bibinfo {author} {\bibfnamefont {Jie}\ \bibnamefont {Lin}}, \bibinfo {author} {\bibfnamefont {Alaa}\ \bibnamefont {Saade}}, \bibinfo {author} {\bibfnamefont {Edan}\ \bibnamefont {Lerner}}, \bibinfo {author} {\bibfnamefont {Alberto}\ \bibnamefont {Rosso}}, \ and\ \bibinfo {author} {\bibfnamefont {Matthieu}\ \bibnamefont {Wyart}},\ }\bibfield  {title} {\enquote {\bibinfo {title} {On the density of shear transformations in amorphous solids},}\ }\href {\doibase 10.1209/0295-5075/105/26003} {\bibfield  {journal} {\bibinfo  {journal} {EPL (Europhysics Letters)}\ }\textbf {\bibinfo {volume} {105}},\ \bibinfo {pages} {26003} (\bibinfo {year} {2014}{\natexlab{a}})}\BibitemShut {NoStop}%
\bibitem [{\citenamefont {M{\" u}ller}\ and\ \citenamefont {Wyart}(2015)}]{Mueller2015}%
  \BibitemOpen
  \bibfield  {author} {\bibinfo {author} {\bibfnamefont {Markus}\ \bibnamefont {M{\" u}ller}}\ and\ \bibinfo {author} {\bibfnamefont {Matthieu}\ \bibnamefont {Wyart}},\ }\bibfield  {title} {\enquote {\bibinfo {title} {Marginal stability in structural, spin, and electron glasses},}\ }\href {\doibase 10.1146/annurev-conmatphys-031214-014614} {\bibfield  {journal} {\bibinfo  {journal} {Annual Review of Condensed Matter Physics}\ }\textbf {\bibinfo {volume} {6}},\ \bibinfo {pages} {177–200} (\bibinfo {year} {2015})}\BibitemShut {NoStop}%
\bibitem [{\citenamefont {Lin}\ \emph {et~al.}(2015)\citenamefont {Lin}, \citenamefont {Gueudr{\' e}}, \citenamefont {Rosso},\ and\ \citenamefont {Wyart}}]{Lin2015}%
  \BibitemOpen
  \bibfield  {author} {\bibinfo {author} {\bibfnamefont {Jie}\ \bibnamefont {Lin}}, \bibinfo {author} {\bibfnamefont {Thomas}\ \bibnamefont {Gueudr{\' e}}}, \bibinfo {author} {\bibfnamefont {Alberto}\ \bibnamefont {Rosso}}, \ and\ \bibinfo {author} {\bibfnamefont {Matthieu}\ \bibnamefont {Wyart}},\ }\bibfield  {title} {\enquote {\bibinfo {title} {Criticality in the approach to failure in amorphous solids},}\ }\href {\doibase 10.1103/PhysRevLett.115.168001} {\bibfield  {journal} {\bibinfo  {journal} {Physical Review Letters}\ }\textbf {\bibinfo {volume} {115}} (\bibinfo {year} {2015}),\ 10.1103/PhysRevLett.115.168001}\BibitemShut {NoStop}%
\bibitem [{\citenamefont {Lin}\ and\ \citenamefont {Wyart}(2016)}]{Lin2016a}%
  \BibitemOpen
  \bibfield  {author} {\bibinfo {author} {\bibfnamefont {Jie}\ \bibnamefont {Lin}}\ and\ \bibinfo {author} {\bibfnamefont {Matthieu}\ \bibnamefont {Wyart}},\ }\bibfield  {title} {\enquote {\bibinfo {title} {Mean-field description of plastic flow in amorphous solids},}\ }\href {\doibase 10.1103/PhysRevX.6.011005} {\bibfield  {journal} {\bibinfo  {journal} {Physical Review X}\ }\textbf {\bibinfo {volume} {6}} (\bibinfo {year} {2016}),\ 10.1103/PhysRevX.6.011005}\BibitemShut {NoStop}%
\bibitem [{\citenamefont {Tahaei}\ \emph {et~al.}(2025)\citenamefont {Tahaei}, \citenamefont {Piscitello-G\'omez}, \citenamefont {Suganthan}, \citenamefont {Cwikla}, \citenamefont {Fuhrmann}, \citenamefont {Dye},\ and\ \citenamefont {Popovi\ifmmode~\acute{c}\else \'{c}\fi{}}}]{Tahaei2025}%
  \BibitemOpen
  \bibfield  {author} {\bibinfo {author} {\bibfnamefont {Ali}\ \bibnamefont {Tahaei}}, \bibinfo {author} {\bibfnamefont {Romina}\ \bibnamefont {Piscitello-G\'omez}}, \bibinfo {author} {\bibfnamefont {S.}~\bibnamefont {Suganthan}}, \bibinfo {author} {\bibfnamefont {Greta}\ \bibnamefont {Cwikla}}, \bibinfo {author} {\bibfnamefont {Jana~F.}\ \bibnamefont {Fuhrmann}}, \bibinfo {author} {\bibfnamefont {Natalie~A.}\ \bibnamefont {Dye}}, \ and\ \bibinfo {author} {\bibfnamefont {Marko}\ \bibnamefont {Popovi\ifmmode~\acute{c}\else \'{c}\fi{}}},\ }\bibfield  {title} {\enquote {\bibinfo {title} {Cell divisions imprint long lasting elastic strain fields in epithelial tissues},}\ }\href {\doibase 10.1103/lh3v-v2c9} {\bibfield  {journal} {\bibinfo  {journal} {PRX Life}\ }\textbf {\bibinfo {volume} {3}},\ \bibinfo {pages} {043008} (\bibinfo {year} {2025})}\BibitemShut {NoStop}%
\bibitem [{\citenamefont {Baret}\ \emph {et~al.}(2002)\citenamefont {Baret}, \citenamefont {Vandembroucq},\ and\ \citenamefont {Roux}}]{Baret2002}%
  \BibitemOpen
  \bibfield  {author} {\bibinfo {author} {\bibfnamefont {J.-C.}\ \bibnamefont {Baret}}, \bibinfo {author} {\bibfnamefont {D.}~\bibnamefont {Vandembroucq}}, \ and\ \bibinfo {author} {\bibfnamefont {S.}~\bibnamefont {Roux}},\ }\bibfield  {title} {\enquote {\bibinfo {title} {Extremal model for amorphous media plasticity},}\ }\href {\doibase 10.1103/PhysRevLett.89.195506} {\bibfield  {journal} {\bibinfo  {journal} {Phys. Rev. Lett.}\ }\textbf {\bibinfo {volume} {89}},\ \bibinfo {pages} {195506} (\bibinfo {year} {2002})}\BibitemShut {NoStop}%
\bibitem [{\citenamefont {Jagla}(2007)}]{Jagla2007}%
  \BibitemOpen
  \bibfield  {author} {\bibinfo {author} {\bibfnamefont {E.A.}\ \bibnamefont {Jagla}},\ }\bibfield  {title} {\enquote {\bibinfo {title} {Strain localization driven by structural relaxation in sheared amorphous solids},}\ }\href {\doibase 10.1103/PhysRevE.76.046119} {\bibfield  {journal} {\bibinfo  {journal} {Phys. Rev. E}\ }\textbf {\bibinfo {volume} {76}},\ \bibinfo {pages} {046119} (\bibinfo {year} {2007})}\BibitemShut {NoStop}%
\bibitem [{\citenamefont {Jagla}(2020)}]{Jagla2020}%
  \BibitemOpen
  \bibfield  {author} {\bibinfo {author} {\bibfnamefont {E.~A.}\ \bibnamefont {Jagla}},\ }\bibfield  {title} {\enquote {\bibinfo {title} {Tensorial description of the plasticity of amorphous composites},}\ }\href {\doibase 10.1103/PhysRevE.101.043004} {\bibfield  {journal} {\bibinfo  {journal} {Physical Review E}\ }\textbf {\bibinfo {volume} {101}},\ \bibinfo {pages} {043004} (\bibinfo {year} {2020})}\BibitemShut {NoStop}%
\bibitem [{\citenamefont {Argon}(1979)}]{Argon1979}%
  \BibitemOpen
  \bibfield  {author} {\bibinfo {author} {\bibfnamefont {A.S}\ \bibnamefont {Argon}},\ }\bibfield  {title} {\enquote {\bibinfo {title} {Plastic deformation in metallic glasses},}\ }\href {\doibase 10.1016/0001-6160(79)90055-5} {\bibfield  {journal} {\bibinfo  {journal} {Acta Metallurgica}\ }\textbf {\bibinfo {volume} {27}},\ \bibinfo {pages} {47–58} (\bibinfo {year} {1979})}\BibitemShut {NoStop}%
\bibitem [{\citenamefont {Lemaitre}\ and\ \citenamefont {Caroli}(2009)}]{Lemaitre2009}%
  \BibitemOpen
  \bibfield  {author} {\bibinfo {author} {\bibfnamefont {Ana{\" e}l}\ \bibnamefont {Lemaitre}}\ and\ \bibinfo {author} {\bibfnamefont {Christiane}\ \bibnamefont {Caroli}},\ }\bibfield  {title} {\enquote {\bibinfo {title} {Rate-dependent avalanche size in athermally sheared amorphous solids},}\ }\href {\doibase 10.1103/PhysRevLett.103.065501} {\bibfield  {journal} {\bibinfo  {journal} {Physical Review Letters}\ }\textbf {\bibinfo {volume} {103}},\ \bibinfo {pages} {065501} (\bibinfo {year} {2009})}\BibitemShut {NoStop}%
\bibitem [{\citenamefont {Farhadifar}\ \emph {et~al.}(2007)\citenamefont {Farhadifar}, \citenamefont {R{\" o}per}, \citenamefont {Aigouy}, \citenamefont {Eaton},\ and\ \citenamefont {J{\" u}licher}}]{Farhadifar2007}%
  \BibitemOpen
  \bibfield  {author} {\bibinfo {author} {\bibfnamefont {Reza}\ \bibnamefont {Farhadifar}}, \bibinfo {author} {\bibfnamefont {Jens-Christian}\ \bibnamefont {R{\" o}per}}, \bibinfo {author} {\bibfnamefont {Benoit}\ \bibnamefont {Aigouy}}, \bibinfo {author} {\bibfnamefont {Suzanne}\ \bibnamefont {Eaton}}, \ and\ \bibinfo {author} {\bibfnamefont {Frank}\ \bibnamefont {J{\" u}licher}},\ }\bibfield  {title} {\enquote {\bibinfo {title} {The influence of cell mechanics, cell-cell interactions, and proliferation on epithelial packing},}\ }\href {\doibase 10.1016/j.cub.2007.11.049} {\bibfield  {journal} {\bibinfo  {journal} {Current Biology}\ }\textbf {\bibinfo {volume} {17}},\ \bibinfo {pages} {2095–2104} (\bibinfo {year} {2007})}\BibitemShut {NoStop}%
\bibitem [{\citenamefont {Fletcher}\ \emph {et~al.}(2014)\citenamefont {Fletcher}, \citenamefont {Osterfield}, \citenamefont {Baker},\ and\ \citenamefont {Shvartsman}}]{Fletcher2014}%
  \BibitemOpen
  \bibfield  {author} {\bibinfo {author} {\bibfnamefont {Alexander~G.}\ \bibnamefont {Fletcher}}, \bibinfo {author} {\bibfnamefont {Miriam}\ \bibnamefont {Osterfield}}, \bibinfo {author} {\bibfnamefont {Ruth~E.}\ \bibnamefont {Baker}}, \ and\ \bibinfo {author} {\bibfnamefont {Stanislav~Y.}\ \bibnamefont {Shvartsman}},\ }\bibfield  {title} {\enquote {\bibinfo {title} {Vertex models of epithelial morphogenesis},}\ }\href {\doibase 10.1016/j.bpj.2013.11.4498} {\bibfield  {journal} {\bibinfo  {journal} {Biophysical Journal}\ }\textbf {\bibinfo {volume} {106}},\ \bibinfo {pages} {2291–2304} (\bibinfo {year} {2014})}\BibitemShut {NoStop}%
\bibitem [{\citenamefont {Eshelby}(1957)}]{Eshelby1957}%
  \BibitemOpen
  \bibfield  {author} {\bibinfo {author} {\bibfnamefont {John~Douglas}\ \bibnamefont {Eshelby}},\ }\bibfield  {title} {\enquote {\bibinfo {title} {The determination of the elastic field of an ellipsoidal inclusion, and related problems},}\ }\href@noop {} {\bibfield  {journal} {\bibinfo  {journal} {Proceedings of the royal society of London. Series A. Mathematical and physical sciences}\ }\textbf {\bibinfo {volume} {241}},\ \bibinfo {pages} {376--396} (\bibinfo {year} {1957})}\BibitemShut {NoStop}%
\bibitem [{\citenamefont {Hertwig}(1884)}]{Hertwig1884}%
  \BibitemOpen
  \bibfield  {author} {\bibinfo {author} {\bibfnamefont {O.}~\bibnamefont {Hertwig}},\ }\href@noop {} {\emph {\bibinfo {title} {Das Problem der Befruchtung und der Isotropie des Eies: eine Theorie der Vererbung}}},\ Jenaische Zeitschrift f\"ur Naturwissenschaften\ (\bibinfo  {publisher} {Fischer},\ \bibinfo {year} {1884})\BibitemShut {NoStop}%
\bibitem [{\citenamefont {Bosveld}\ \emph {et~al.}(2016)\citenamefont {Bosveld}, \citenamefont {Markova}, \citenamefont {Guirao}, \citenamefont {Martin}, \citenamefont {Wang}, \citenamefont {Pierre}, \citenamefont {Balakireva}, \citenamefont {Gaugue}, \citenamefont {Ainslie}, \citenamefont {Christophorou}, \citenamefont {Lubensky}, \citenamefont {Minc},\ and\ \citenamefont {Bellaïche}}]{Bosveld2016}%
  \BibitemOpen
  \bibfield  {author} {\bibinfo {author} {\bibfnamefont {Floris}\ \bibnamefont {Bosveld}}, \bibinfo {author} {\bibfnamefont {Olga}\ \bibnamefont {Markova}}, \bibinfo {author} {\bibfnamefont {Boris}\ \bibnamefont {Guirao}}, \bibinfo {author} {\bibfnamefont {Charlotte}\ \bibnamefont {Martin}}, \bibinfo {author} {\bibfnamefont {Zhimin}\ \bibnamefont {Wang}}, \bibinfo {author} {\bibfnamefont {Anaëlle}\ \bibnamefont {Pierre}}, \bibinfo {author} {\bibfnamefont {Maria}\ \bibnamefont {Balakireva}}, \bibinfo {author} {\bibfnamefont {Isabelle}\ \bibnamefont {Gaugue}}, \bibinfo {author} {\bibfnamefont {Anna}\ \bibnamefont {Ainslie}}, \bibinfo {author} {\bibfnamefont {Nicolas}\ \bibnamefont {Christophorou}}, \bibinfo {author} {\bibfnamefont {David~K.}\ \bibnamefont {Lubensky}}, \bibinfo {author} {\bibfnamefont {Nicolas}\ \bibnamefont {Minc}}, \ and\ \bibinfo {author} {\bibfnamefont {Yohanns}\ \bibnamefont {Bellaïche}},\ }\bibfield  {title} {\enquote {\bibinfo {title} {Epithelial tricellular junctions act as interphase cell shape sensors to orient mitosis},}\ }\href {\doibase 10.1038/nature16970} {\bibfield  {journal} {\bibinfo  {journal} {Nature}\ }\textbf {\bibinfo {volume} {530}},\ \bibinfo {pages} {495--498} (\bibinfo {year} {2016})}\BibitemShut {NoStop}%
\bibitem [{\citenamefont {Herschel}\ and\ \citenamefont {Bulkley}(1926)}]{Herschel1926}%
  \BibitemOpen
  \bibfield  {author} {\bibinfo {author} {\bibfnamefont {Winslow~H.}\ \bibnamefont {Herschel}}\ and\ \bibinfo {author} {\bibfnamefont {Ronald}\ \bibnamefont {Bulkley}},\ }\bibfield  {title} {\enquote {\bibinfo {title} {Konsistenzmessungen von gummi-benzoll{\"o}sungen},}\ }\href {\doibase 10.1007/BF01432034} {\bibfield  {journal} {\bibinfo  {journal} {Kolloid-Zeitschrift}\ }\textbf {\bibinfo {volume} {39}},\ \bibinfo {pages} {291--300} (\bibinfo {year} {1926})}\BibitemShut {NoStop}%
\bibitem [{\citenamefont {Karmakar}\ \emph {et~al.}(2010)\citenamefont {Karmakar}, \citenamefont {Lerner},\ and\ \citenamefont {Procaccia}}]{Karmakar2010}%
  \BibitemOpen
  \bibfield  {author} {\bibinfo {author} {\bibfnamefont {Smarajit}\ \bibnamefont {Karmakar}}, \bibinfo {author} {\bibfnamefont {Edan}\ \bibnamefont {Lerner}}, \ and\ \bibinfo {author} {\bibfnamefont {Itamar}\ \bibnamefont {Procaccia}},\ }\bibfield  {title} {\enquote {\bibinfo {title} {Statistical physics of the yielding transition in amorphous solids},}\ }\href {\doibase 10.1103/PhysRevE.82.055103} {\bibfield  {journal} {\bibinfo  {journal} {Physical Review E}\ }\textbf {\bibinfo {volume} {82}} (\bibinfo {year} {2010}),\ 10.1103/PhysRevE.82.055103}\BibitemShut {NoStop}%
\bibitem [{\citenamefont {Lin}\ \emph {et~al.}(2014{\natexlab{b}})\citenamefont {Lin}, \citenamefont {Lerner}, \citenamefont {Rosso},\ and\ \citenamefont {Wyart}}]{Lin2014b}%
  \BibitemOpen
  \bibfield  {author} {\bibinfo {author} {\bibfnamefont {Jie}\ \bibnamefont {Lin}}, \bibinfo {author} {\bibfnamefont {Edan}\ \bibnamefont {Lerner}}, \bibinfo {author} {\bibfnamefont {Alberto}\ \bibnamefont {Rosso}}, \ and\ \bibinfo {author} {\bibfnamefont {Matthieu}\ \bibnamefont {Wyart}},\ }\bibfield  {title} {\enquote {\bibinfo {title} {Scaling description of the yielding transition in soft amorphous solids at zero temperature},}\ }\href {\doibase 10.1073/pnas.1406391111} {\bibfield  {journal} {\bibinfo  {journal} {Proceedings of the National Academy of Sciences}\ }\textbf {\bibinfo {volume} {111}},\ \bibinfo {pages} {14382–14387} (\bibinfo {year} {2014}{\natexlab{b}})}\BibitemShut {NoStop}%
\bibitem [{\citenamefont {H\'ebraud}\ and\ \citenamefont {Lequeux}(1998)}]{Hebraud1998}%
  \BibitemOpen
  \bibfield  {author} {\bibinfo {author} {\bibfnamefont {P.}~\bibnamefont {H\'ebraud}}\ and\ \bibinfo {author} {\bibfnamefont {F.}~\bibnamefont {Lequeux}},\ }\bibfield  {title} {\enquote {\bibinfo {title} {Mode-coupling theory for the pasty rheology of soft glassy materials},}\ }\href {\doibase 10.1103/PhysRevLett.81.2934} {\bibfield  {journal} {\bibinfo  {journal} {Phys. Rev. Lett.}\ }\textbf {\bibinfo {volume} {81}},\ \bibinfo {pages} {2934--2937} (\bibinfo {year} {1998})}\BibitemShut {NoStop}%
\bibitem [{\citenamefont {Agoritsas}\ \emph {et~al.}(2015)\citenamefont {Agoritsas}, \citenamefont {Bertin}, \citenamefont {Martens},\ and\ \citenamefont {Barrat}}]{Agoritsas2015}%
  \BibitemOpen
  \bibfield  {author} {\bibinfo {author} {\bibfnamefont {Elisabeth}\ \bibnamefont {Agoritsas}}, \bibinfo {author} {\bibfnamefont {Eric}\ \bibnamefont {Bertin}}, \bibinfo {author} {\bibfnamefont {Kirsten}\ \bibnamefont {Martens}}, \ and\ \bibinfo {author} {\bibfnamefont {Jean-Louis}\ \bibnamefont {Barrat}},\ }\bibfield  {title} {\enquote {\bibinfo {title} {On the relevance of disorder in athermal amorphous materials under shear},}\ }\href {\doibase 10.1140/epje/i2015-15071-x} {\bibfield  {journal} {\bibinfo  {journal} {The European Physical Journal E}\ }\textbf {\bibinfo {volume} {38}},\ \bibinfo {pages} {71} (\bibinfo {year} {2015})}\BibitemShut {NoStop}%
\bibitem [{\citenamefont {Popovi{\'c}}\ \emph {et~al.}(2021)\citenamefont {Popovi{\'c}}, \citenamefont {de~Geus},\ and\ \citenamefont {Wyart}}]{Popovic2021Thermal}%
  \BibitemOpen
  \bibfield  {author} {\bibinfo {author} {\bibfnamefont {Marko}\ \bibnamefont {Popovi{\'c}}}, \bibinfo {author} {\bibfnamefont {Tom W.~J.}\ \bibnamefont {de~Geus}}, \ and\ \bibinfo {author} {\bibfnamefont {Matthieu}\ \bibnamefont {Wyart}},\ }\bibfield  {title} {\enquote {\bibinfo {title} {Thermally activated flow in models of amorphous solids},}\ }\href {\doibase 10.1103/PhysRevE.104.025010} {\bibfield  {journal} {\bibinfo  {journal} {Physical Review E}\ }\textbf {\bibinfo {volume} {104}},\ \bibinfo {pages} {025010} (\bibinfo {year} {2021})}\BibitemShut {NoStop}%
\bibitem [{\citenamefont {Popović}\ \emph {et~al.}(2021)\citenamefont {Popović}, \citenamefont {Druelle}, \citenamefont {Dye}, \citenamefont {Jülicher},\ and\ \citenamefont {Wyart}}]{Popovic2021a}%
  \BibitemOpen
  \bibfield  {author} {\bibinfo {author} {\bibfnamefont {Marko}\ \bibnamefont {Popović}}, \bibinfo {author} {\bibfnamefont {Valentin}\ \bibnamefont {Druelle}}, \bibinfo {author} {\bibfnamefont {Natalie~A}\ \bibnamefont {Dye}}, \bibinfo {author} {\bibfnamefont {Frank}\ \bibnamefont {Jülicher}}, \ and\ \bibinfo {author} {\bibfnamefont {Matthieu}\ \bibnamefont {Wyart}},\ }\bibfield  {title} {\enquote {\bibinfo {title} {Inferring the flow properties of epithelial tissues from their geometry},}\ }\href {\doibase 10.1088/1367-2630/abcbc7} {\bibfield  {journal} {\bibinfo  {journal} {New Journal of Physics}\ }\textbf {\bibinfo {volume} {23}},\ \bibinfo {pages} {033004} (\bibinfo {year} {2021})}\BibitemShut {NoStop}%
\bibitem [{\citenamefont {Harris}\ \emph {et~al.}(2012)\citenamefont {Harris}, \citenamefont {Peter}, \citenamefont {Bellis}, \citenamefont {Baum}, \citenamefont {Kabla},\ and\ \citenamefont {Charras}}]{Harris2012}%
  \BibitemOpen
  \bibfield  {author} {\bibinfo {author} {\bibfnamefont {A.~R.}\ \bibnamefont {Harris}}, \bibinfo {author} {\bibfnamefont {L.}~\bibnamefont {Peter}}, \bibinfo {author} {\bibfnamefont {J.}~\bibnamefont {Bellis}}, \bibinfo {author} {\bibfnamefont {B.}~\bibnamefont {Baum}}, \bibinfo {author} {\bibfnamefont {A.~J.}\ \bibnamefont {Kabla}}, \ and\ \bibinfo {author} {\bibfnamefont {G.~T.}\ \bibnamefont {Charras}},\ }\bibfield  {title} {\enquote {\bibinfo {title} {Characterizing the mechanics of cultured cell monolayers},}\ }\href {\doibase 10.1073/pnas.1213301109} {\bibfield  {journal} {\bibinfo  {journal} {Proceedings of the National Academy of Sciences}\ }\textbf {\bibinfo {volume} {109}},\ \bibinfo {pages} {16449–16454} (\bibinfo {year} {2012})}\BibitemShut {NoStop}%
\bibitem [{\citenamefont {Schotz}\ \emph {et~al.}(2013)\citenamefont {Schotz}, \citenamefont {Lanio}, \citenamefont {Talbot},\ and\ \citenamefont {Manning}}]{Schotz2013}%
  \BibitemOpen
  \bibfield  {author} {\bibinfo {author} {\bibfnamefont {E.-M.}\ \bibnamefont {Schotz}}, \bibinfo {author} {\bibfnamefont {M.}~\bibnamefont {Lanio}}, \bibinfo {author} {\bibfnamefont {J.~A.}\ \bibnamefont {Talbot}}, \ and\ \bibinfo {author} {\bibfnamefont {M.~L.}\ \bibnamefont {Manning}},\ }\bibfield  {title} {\enquote {\bibinfo {title} {Glassy dynamics in three-dimensional embryonic tissues},}\ }\href {\doibase 10.1098/rsif.2013.0726} {\bibfield  {journal} {\bibinfo  {journal} {Journal of The Royal Society Interface}\ }\textbf {\bibinfo {volume} {10}},\ \bibinfo {pages} {20130726–20130726} (\bibinfo {year} {2013})}\BibitemShut {NoStop}%
\bibitem [{\citenamefont {Curran}\ \emph {et~al.}(2017)\citenamefont {Curran}, \citenamefont {Strandkvist}, \citenamefont {Bathmann}, \citenamefont {de~Gennes}, \citenamefont {Kabla}, \citenamefont {Salbreux},\ and\ \citenamefont {Baum}}]{Curran2017}%
  \BibitemOpen
  \bibfield  {author} {\bibinfo {author} {\bibfnamefont {Scott}\ \bibnamefont {Curran}}, \bibinfo {author} {\bibfnamefont {Charlotte}\ \bibnamefont {Strandkvist}}, \bibinfo {author} {\bibfnamefont {Jasper}\ \bibnamefont {Bathmann}}, \bibinfo {author} {\bibfnamefont {Marc}\ \bibnamefont {de~Gennes}}, \bibinfo {author} {\bibfnamefont {Alexandre}\ \bibnamefont {Kabla}}, \bibinfo {author} {\bibfnamefont {Guillaume}\ \bibnamefont {Salbreux}}, \ and\ \bibinfo {author} {\bibfnamefont {Buzz}\ \bibnamefont {Baum}},\ }\bibfield  {title} {\enquote {\bibinfo {title} {Myosin ii controls junction fluctuations to guide epithelial tissue ordering},}\ }\href {\doibase 10.1016/j.devcel.2017.09.018} {\bibfield  {journal} {\bibinfo  {journal} {Developmental Cell}\ }\textbf {\bibinfo {volume} {43}},\ \bibinfo {pages} {480--492.e6} (\bibinfo {year} {2017})},\ \bibinfo {note} {publisher: Elsevier}\BibitemShut {NoStop}%
\bibitem [{\citenamefont {Bi}\ \emph {et~al.}(2016)\citenamefont {Bi}, \citenamefont {Yang}, \citenamefont {Marchetti},\ and\ \citenamefont {Manning}}]{Bi2016}%
  \BibitemOpen
  \bibfield  {author} {\bibinfo {author} {\bibfnamefont {Dapeng}\ \bibnamefont {Bi}}, \bibinfo {author} {\bibfnamefont {Xingbo}\ \bibnamefont {Yang}}, \bibinfo {author} {\bibfnamefont {M.~Cristina}\ \bibnamefont {Marchetti}}, \ and\ \bibinfo {author} {\bibfnamefont {M.~Lisa}\ \bibnamefont {Manning}},\ }\bibfield  {title} {\enquote {\bibinfo {title} {Motility-driven glass and jamming transitions in biological tissues},}\ }\href {\doibase 10.1103/PhysRevX.6.021011} {\bibfield  {journal} {\bibinfo  {journal} {Physical Review X}\ }\textbf {\bibinfo {volume} {6}} (\bibinfo {year} {2016}),\ 10.1103/PhysRevX.6.021011}\BibitemShut {NoStop}%
\bibitem [{\citenamefont {Yamamoto}\ \emph {et~al.}(2022)\citenamefont {Yamamoto}, \citenamefont {Sussman}, \citenamefont {Shibata},\ and\ \citenamefont {Manning}}]{Yamamoto2022}%
  \BibitemOpen
  \bibfield  {author} {\bibinfo {author} {\bibfnamefont {Takaki}\ \bibnamefont {Yamamoto}}, \bibinfo {author} {\bibfnamefont {Daniel~M.}\ \bibnamefont {Sussman}}, \bibinfo {author} {\bibfnamefont {Tatsuo}\ \bibnamefont {Shibata}}, \ and\ \bibinfo {author} {\bibfnamefont {M.~Lisa}\ \bibnamefont {Manning}},\ }\bibfield  {title} {\enquote {\bibinfo {title} {Non-monotonic fluidization generated by fluctuating edge tensions in confluent tissues},}\ }\href {\doibase 10.1039/D0SM01559H} {\bibfield  {journal} {\bibinfo  {journal} {Soft Matter}\ }\textbf {\bibinfo {volume} {18}},\ \bibinfo {pages} {2168--2175} (\bibinfo {year} {2022})}\BibitemShut {NoStop}%
\bibitem [{Note1()}]{Note1}%
  \BibitemOpen
  \bibinfo {note} {We only consider the pure shear components of strain and stress, so in the following we omit the 'pure shear' when denoting strain and stress to keep the notation short.}\BibitemShut {Stop}%
\end{thebibliography}%

%\newpage

\appendix

\newpage
\section{Implementation of cell divisions in the elasto-plastic model}\label{appendix:division_implementation}
Here, we present a more detailed discussion on the implementation of cell divisions in the simulations.
We first consider randomly oriented divisions where the axis of division is randomly oriented, independent of the stress in the block.
In 2d system, the pure shear plastic strain\footnote{We only consider the pure shear components of strain and stress, so in the following we omit the 'pure shear' when denoting strain and stress to keep the notation short.} accumulated by a division is
\begin{align*}
    \Delta  \epsilon {}_{xx}^{pl} = \zeta \cos[\Phi],
    \\
    \Delta  \epsilon {}_{xy}^{pl} = \zeta \sin[\Phi].
\end{align*}
with uniformly distributed random numbers $\zeta \in [0, \Delta_{\rm{div}}]$ and $\Phi \in [0,2\pi]$.
In the simulations performed in this work text, we use  division magnitude $\Delta_{\rm{div}}=1$, so that the accumulated plastic strain $\Delta \mathbf{{\epsilon}}$ has a variance of $\eta^2_{\rm{div}}=1/6$. %The precise value of $\eta_{\rm{div}}$ mainly sets the magnitude of the active kick, while the qualitative distinction between randomly oriented and stress-relaxing divisions is controlled by the orientation and resetting rules.

The second type of cell division we consider are stress relaxing divisions, after which the local shear stress is fully relaxed
\begin{align*}
    \Delta \epsilon {}_{xx}^{pl} &= \sigma {}_{xx}(\vec r_0),
    \\
    \Delta \epsilon {}_{xy}^{pl} &= \sigma {}_{xy}(\vec r_0),
\end{align*}
where $\vec r_0$ is the position of the block in which the division occurs.

Finally, for stress-oriented divisions, the plastic strain is taken to be aligned with the local stress tensor, while its magnitude is drawn randomly from a bounded exponential probability distribution function. Then, 
\begin{align}
    \Delta \epsilon {}_{xx}^{pl} &= \frac{ \sigma {}_{xx}(\vec r_0)}{{\sigma} (\vec{r}_0)} \zeta,
    \\
    \Delta \epsilon {}_{xy}^{pl} &= \frac{ \sigma {}_{xy}(\vec r_0)}{{\sigma} (\vec{r}_0)} \zeta,
\end{align}
where $\zeta$ is a random variable drawn from a bounded exponential probability distribution function. In this work, we use the distribution of $\zeta$,
\begin{align}
    P(\zeta) = A e^{-\zeta/\sigma_Y}, \qquad \zeta \in [\sigma_Y, 2\sigma_Y],
\end{align}
where $A$ is a normalization constant and $\sigma_Y$ is the local yield stress. 
With this choice, the cell divisions contribute directly to the imposed shear on average, unlike randomly oriented divisions, but without fully resetting the local stress as with the SRD divisions, see Appendix \ref{appendix:division_el_balance}.

\section{Estimation of the pseudogap exponent from the simulations}
\begin{figure}
    \centering
    \includegraphics[width=0.99\linewidth]{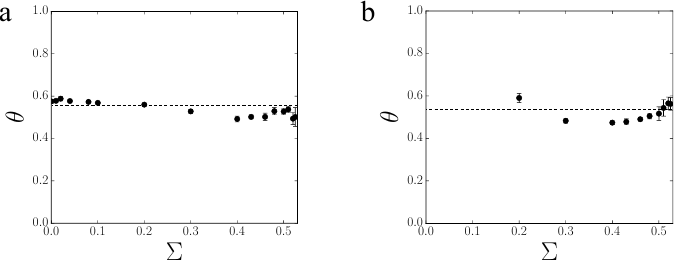}
    \caption{Pseudogap exponent $\theta$ as a function of the applied shear stress $\Sigma$ for (a) randomly oriented and (b) stress-relaxing divisions. For each $\Sigma$, $\theta$ is extracted from a power-law fit of the low-$x$ tail of $P(x)$ (Fig.~\ref{fig:EPM_division_pseudogap}) over $10^{-3}\leq x < 3\times 10^{-2}$. The dashed line indicates the average over $\Sigma$, yielding $\theta=0.55\pm 0.03$ in (a) and $\theta=0.52\pm 0.04$ in (b), with no clear systematic dependence on $\Sigma$.}
    \label{fig:theta_simulations}
\end{figure}

As shown in Fig.~\ref{fig:EPM_division_pseudogap}, the steady-state distribution of local stabilities exhibits a pseudogap, $P(x)\sim x^\theta$ at small $x$. To estimate $\theta$, we fit a power law over the window $10^{-3}\leq x < 3\times 10^{-2}$ for each simulation, and plot the resulting exponent as a function of the applied shear stress $\Sigma$ in Fig.~\ref{fig:theta_simulations}. Within our numerical uncertainty, we find no systematic dependence of $\theta$ on $\Sigma$; we therefore average over $\Sigma$ and obtain $\theta=0.55\pm 0.03$ for randomly oriented divisions and $\theta=0.52\pm 0.04$ for stress-relaxing divisions, both consistent with $\theta \approx 0.57$ often reported in litreture \cite{Lin2014a,Nicolas2018}. For stress-relaxing divisions at low $\Sigma$, no reliable value of $\theta$ can be extracted because low local stabilities are strongly depleted, see the main text and Appendix~\ref{appendix:Hebraud-Lequeux} for more discussions.

\section{H\'ebraud-Lequeux model with cell divisions}\label{appendix:Hebraud-Lequeux}

In this Appendix, we present the analysis of H\'ebraud-Lequeux model \cite{Hebraud1998,Agoritsas2015} at $\Sigma= 0$. The point is to show analytically how stress-relaxing divisions suppress the distribution $P(x)$ but without opening a true gap.

Previously, randomly oriented divisions were considered \cite{Matoz-Fernandez2017}, but here we need to additionally account for the stress relaxation due to stress-relaxing divisions. We consider the stationary H\'ebraud-Lequeux (HL) equation at $\Sigma=0$, with local stress $\sigma$ measured in units of the yield stress $\sigma_Y=1$. Mechanical noise inducing the stress diffusion stems from cell rearrangement contributing $\alpha \Gamma$ to the diffusion coefficient, where $\Gamma$ is the rate of cell rearrangements. Cell divisions further contribute $D_0 k_d$. The HL equation then reads
\begin{align}
 0& =
    \left(\alpha \Gamma + D_0 k_d\right)\partial_\sigma^2 P(\sigma)
    + \Gamma\delta(\sigma)\\
    &- \frac{1}{\tau_0}\Theta(|\sigma|-1)P(\sigma)
    + \phi k_d\left[\delta(\sigma)-P(\sigma)\right].
\end{align}
where 
$\phi = 0 \text{ for ROD}$ and $\phi = 1  \text{ for SRD}.$ so that randomly oriented divisions contribute only to the diffusive spreading of stress. Stress-relaxing divisions additionally reset the stress of the dividing block to $0$.

In the quasistatic division limit, it is convenient to replace the time by the count of cell division events, $n=k_d t$, and define the dimensionless activity $\gamma=\Gamma/k_d$. The total diffusion constant then reads
\begin{align}
    D_m=\alpha \gamma + D_0.
\end{align}

\subsection{Randomly oriented divisions}

For randomly oriented divisions, the stationary zero-stress equation inside the stable interval $|\sigma|<1$ reduces to diffusion with reinjection at $\sigma=0$. The symmetric stationary solution is
\begin{align}
P(\sigma)=
\begin{cases}
\dfrac{1}{2D_m}\sigma+\dfrac{1}{2D_m}, & -1\leq \sigma \leq 0,\\[4pt]
-\dfrac{1}{2D_m}\sigma+\dfrac{1}{2D_m}, & 0\leq \sigma \leq 1,
\end{cases}
\end{align}
which is the standard mean-field pseudogap H\'ebraud-Lequeux model. Defining the distance to instability as $x=1-\sigma$ for $\sigma>0$, one obtains
\begin{align}
    P(x)\sim A_{\rm ROD}x, \qquad A_{\rm ROD}=\frac{1}{2D_m}.
\end{align}
Thus, randomly oriented divisions preserve a linear pseudogap with a prefactor that remains finite in the low-noise regime.

\subsection{Stress-relaxing divisions}

For stress-relaxing divisions, the stationary equation gains an additional resetting term. At zero imposed stress, the distribution in the stable interval obeys
\begin{align}
    D_m\,\partial_\sigma^2 P(\sigma) + (\gamma+1)\delta(\sigma) - P(\sigma)=0,
    \qquad |\sigma|<1,
    \label{eq:HL_zero_stress_relaxing}
\end{align}
with boundary conditions $P(\pm 1)=0$. Away from $\sigma=0$ the solution is exponential, and matching the two branches at the origin gives
\onecolumngrid

\begin{align}
P(\sigma)=
\begin{cases}
\dfrac{\gamma+1}{2\sinh(1/\sqrt{D_m})}\left[e^{(\sigma+1)/\sqrt{D_m}}-e^{-(\sigma+1)/\sqrt{D_m}}\right], & -1<\sigma<0,\\[6pt]
\dfrac{\gamma+1}{2\sinh(1/\sqrt{D_m})}\left[e^{(1-\sigma)/\sqrt{D_m}}-e^{-(1-\sigma)/\sqrt{D_m  }}\right], & 0<\sigma<1.
\end{cases}
\label{eq:HL_relaxing_solution_zero_stress}
\end{align}
\twocolumngrid

Expanding near the yield threshold, $\sigma=1-x$ with $x\ll 1$, yields again a linear pseudogap,
\begin{align}
    P(x)\sim A_{\rm SRD}x,
\end{align}
but now with prefactor
\begin{align}
    A_{\rm SRD}=\frac{\gamma+1}{\sqrt{D_m}\sinh(1/\sqrt{D_m})}
    \sim \frac{e^{-1/\sqrt{D_m}}}{D_m}
    \qquad (D_m\ll 1).
\end{align}
The main result is therefore not a change of the pseudogap exponent, but the prefactor, which becomes smaller in the presence of stress-relaxing divisions as we discuss below.

\subsection{Suppression of available plasticity}

The meaning of this result becomes transparent by estimating the typical lowest $x$ in a system of $N$ blocks from
\begin{align}
    \frac{1}{N}\sim \int_0^{x_{\min}} P(x)\,dx.
\end{align}
For a linear pseudogap this gives $x_{\min}\sim (A N)^{-1/2}$. Using the prefactors determined above, we obtain
\begin{align}
    x_{\min}^{\rm ROD}\sim \sqrt{\frac{2D_m}{N}},
    \qquad
    x_{\min}^{\rm SRD}\sim \sqrt{\frac{\sqrt{D}\sinh(1/\sqrt{D_m})}{(\gamma+1)N}}.
\end{align}
In the low-noise limit, this gives
\begin{align}
    x_{\min}^{\rm SRD}\sim \sqrt{\frac{D_m e^{1/\sqrt{D_m}}}{N}},
    \qquad (D_m\ll 1),
\end{align}
where we omitted factors of order unity. Because $x_{\min}^{\rm SRD}$ is larger than $x_{\min}^{\rm ROD}$ in the low-noise regime, the weakest block is farther from instability for stress-relaxing divisions. This is precisely the sense in which available plasticity is suppressed: the system remains  marginally stable, since $P(x)$ still vanishes only at $x=0$, but the number of blocks close enough to yield and participate in cell rearrangement avalanches is drastically reduced. This zero-stress mean-field picture helps us understand the results of the two-dimensional system discussed in the main text.

For the simulations shown in the main text, the division magnitude is $\Delta_{\rm di}=1$. A rough estimate of the corresponding mean-field noise can be obtained by projecting a randomly oriented stress contribution with amplitude drawn from $A\in[0,\Delta_{\rm di}]$ onto one stress component:
\begin{align}
    \langle \delta\sigma^2\rangle \sim \langle A^2\rangle\langle \cos^2\Phi\rangle
    = \frac{\Delta_{\rm di}^2}{3}\frac{1}{2}
    = \frac{1}{6}.
\end{align}
If $D_0\sim \langle \delta\sigma^2\rangle/2$, this gives $D_0\sim 1/12$ and therefore $1/\sqrt{D_0}\sim 3.5$. Using the full kick magnitude instead gives $D_0\sim 1/6$ and $1/\sqrt{D_0}\sim 2.5$. Thus, for $\Delta_{\rm di}=1$ one expects $1/\sqrt{D_m}$ to be of order $2-4$, and the exponential enhancement $e^{1/\sqrt{D_m}}$ to be of order $10-30$, unless rearrangement-induced mechanical noise $\alpha\gamma$ substantially increases $D_m$.

\section{Elastic energy balance in EPM} \label{appendix:elasticEnergy}
We use an elastic-energy balance to relate the number of cell rearrangements to the energy injected by cell divisions. The dimensionless elastic energy is
\begin{align}
    E = \frac{1}{2}\braket{\bm{\sigma}|\bm{\sigma}},
\end{align}
where $\ket{\bm{\sigma}} = ({\sigma}_{xx,1},{\sigma}_{xy,1},\cdots,{\sigma}_{xx,N},{\sigma}_{xy,N})$ is a $2N$ dimensional vector that contains the stress components of all blocks, i.e. $\sigma_{xx,i}$ and $\sigma_{xy,i}$ for $i=1,\dots,N$. We consider a fixed imposed shear stress $\Sigma\equiv\Sigma_{xy}$ and ask how a single event changes the elastic energy. Given a stress change $\ket{\bm{\sigma}} \to \ket{\bm{\sigma}}+\ket{\delta \bm{\sigma}}$, the corresponding energy change is
\begin{align}
    \delta E = \Braket{\bm{\sigma}|\delta \bm{\sigma}} + \frac{1}{2} \braket{\delta\bm{\sigma} | \delta\bm{\sigma} }.
\end{align}

The stress change is caused by either a cell rearrangement or a cell division, jointly denoted as plastic events (passive and active, respectively). The change of stress due to a plastic event $m$, can be effectively represented by a force dipole contained in one block $\ket{\bm{d}_m}= (0,\cdots,0,{d}_{xx,i_m},{d}_{xy,i_m},0\cdots,0)$ in the block $i_m$. The corresponding stress vector change is given by elastic propagation: $\ket{\bm{\sigma}} = \hat{\bm{G}}\ket{\bm{d}_m}$ where $\hat{\bm{G}}$ is the elastic propagator. Keeping the leading terms in $N$, the elastic energy change is
\begin{align}
    \delta E_m = \braket{\bm{\sigma}|{\bm{d}_m}} - \braket{\bm{\sigma}|\hat{\bm{G}}|{\bm{d}_m}} -& {\Sigma}\braket{\bm{u}_{xy}|{\bm{d}_m}} + \nonumber \\
    &(\tfrac{1}{2}+D)\braket{{\bm{d}_m}|{\bm{d}_m}}, \label{eq:EB_discrete_Eevolve}
\end{align}
and $\ket{\bm{u}_{xy}}=(0,1,0,1,\dots)$ projects onto the imposed-stress direction and $\hat{\bm{G}}\ket{\bm{u}_{xy}}=0$ for a stress controlled ensemble. The constant $D$ is set by the kernel via $2D=(\hat{\bm{G}}^2)_{\alpha\alpha}$.

Finally, we denote by $\mathbf{E}[\cdots]$ an average over an ensemble of events in the steady state. Without divisions, for $\Sigma\ge \Sigma_c$, the elastic energy remains constant, so $\mathbf{E}[\delta E_m^{\mathrm{re}}]=0$ for cell rearrangements. In other words, the energy provided by imposed stress is balanced by the energy dissipated through rearrangements. Specially, at ${\Sigma}=\Sigma_c$, we find the yield sterss of the material as, 
\begin{align}
    \Sigma_c = \frac{\mathbf{E}[{\sigma}^2_{i_m}](\frac{1}{2}-D) - \mathbf{E}[\braket{\bm{\sigma}|\hat{\bm{G}}|{\bm{d}_m}}]}{\mathbf{E}[{\sigma}_{xy,i_m}]}
\end{align}
where ${\sigma}_{i_m}^2 = {\sigma}_{xx,i_m}^2 + {\sigma}_{xy,i_m}^2$ is the shear stress of the yielding block.

For $\Sigma<\Sigma_c$ cell rearrangements dissipate elastic energy on average ($\mathbf{E}[\delta E_m^{\mathrm{re}}]<0$), and the dynamics eventually arrests. More specifically, for ${\Sigma} \to \Sigma_c^-$ one finds
\begin{align}
    \mathbf{E}[\delta E_m^{\mathrm{re}}] &= -\mathbf{E}[{\sigma}_{i_m}^2](\frac{1}{2}-D) +{\Sigma} \mathbf{E}[{\sigma}_{i_m}] - \mathbf{E}[\braket{\bm{\sigma}|\hat{\bm{G}}|{\bm{d}_m}}] \nonumber \\
    &\approx -(\Sigma_c - {\Sigma})\mathbf{E}[{\sigma}_{i_m}] + \mathcal{O}[(\Sigma_c - \Sigma)^2]
\end{align}
where we have used $\mathbf{E}[\ket{\bm{\sigma}}_{\Sigma_c} - \ket{\bm{\sigma}}_{{\Sigma}}] \approx (\Sigma_c - {\Sigma}) \ket{\bm{u}_{xy}}$ and only kept the leading terms in $\Sigma_c - {\Sigma}$.

\section{The energy injection by cell divisions controls the avalanche size}\label{appendix:division_el_balance}
Below $\Sigma_c$, mechanically triggered rearrangements dissipate elastic energy. In the steady state, this dissipation must be compensated for by the energy injected by cell divisions. Since, by definition of avalanche size, each cell division is followed on average by $\langle S\rangle$ rearrangements, the elastic energy balance gives
\begin{align}
    \langle S \rangle \mathbf{E}[\delta E_m^{\mathrm{re}}] + \mathbf{E}[\delta E_m^{\mathrm{div}}] = 0. \label{eq:appendix_general_elasitc_E_div}
\end{align}
Therefore, if $\mathbf{E}[\delta E_m^{\mathrm{div}}]$ remains constant close $\Sigma_c$, we generally expect $\langle S \rangle \sim [\Sigma_c - {\Sigma}]^{-1}$ as observed in simulations, see Fig.~\ref{fig:EPM_division_avalanches}. In this section, we calculate the elastic energy injection of cell divisions for the division types defined in the main text and argue about their results in more detail.

\subsubsection{Randomly-oriented divisions}
The force dipole generated by a randomly-oriented division has zero mean, $\mathbf{E}[\ket{\bm{d}}_m]=0$, and is uncorrelated between divisions. From Eq.~\ref{eq:EB_discrete_Eevolve}, this yields
\begin{align}
    \mathbf{E}[\delta E_m^{\mathrm{div}}] \approx \left(\frac{1}{2} + D\right)\eta_{\text{div}}^2, \label{eq:elastic_E_random_div}
\end{align}
where $\eta_{\text{div}}^2$ is the variance of the strain imposed by divisions. Notably, the elastic energy injected by randomly-oriented divisions is always positive and independent of the current system state $\ket{\bm{\sigma}}$.

Inserting Eq.~\ref{eq:elastic_E_random_div} into Eq.~\ref{eq:appendix_general_elasitc_E_div}, we find the average avalanche size,
\begin{align}
    \langle S \rangle = \frac{\left(\frac{1}{2}+D\right)\eta_{\text{div}}^2}{\Sigma_c - {\Sigma}}, \label{eq:avalancehSize_random_theory}
\end{align}
which provides the $\langle S\rangle\sim(\Sigma_c-\Sigma)^{-1}$ scaling used in the main text. Furthermore, the prefactor of this scaling is directly related to the variance of the division-induced strain, which can be measured in simulations. This prediction is in good agreement with our numerical measurements, as shown in Fig.~\ref{fig:EPM_division_avalanches} (e).

\subsubsection{Stress-relaxing divisions}
Similar to a cell rearrangement, a stress-relaxing division relaxes local stress to zero. Unlike cell rearrangements, the cell divisions are not limited to mechanically unstable regions, and all blocks can divide and relax their stress. This modifies the average stress relaxation to $\textbf{E}[{\sigma}_{xy,i_m}]={\Sigma}$. Therefore, each division changes the elastic energy as,
\begin{align}
    \mathbf{E}[\delta E_m^{{\text{div}}}] = -(\frac{1}{2}-D)\mathbf{E}[{\sigma}^2]+{\Sigma}^2. \label{eq:elastic_E_relaxing_div}
\end{align}
This predicts that the individual stress-relaxing divisions can inject or reduce the elastic energy of the tissue, depending on the first and second moments of stress distribution. 
Combining Eq.~\ref{eq:elastic_E_relaxing_div} with the balance condition in Eq.~\ref{eq:appendix_general_elasitc_E_div} yields a quantitative prediction for $\langle S\rangle$ as shown in Fig.~\ref{fig:EPM_division_avalanches} (f). In particular, near yielding $\mathbf{E}[\delta E_m^{\mathrm{div}}]$ remains finite, which is enough to recover $\langle S\rangle\sim(\Sigma_c-\Sigma)^{-1}$. At low imposed stress, the first term in Eq.~\ref{eq:elastic_E_relaxing_div} can dominate, so divisions remove elastic energy on average, narrowing the stress distribution and strongly suppressing mechanically triggered avalanches. This is consistent with our numerical observations that stress-relaxing divisions lead to much smaller avalanches than randomly-oriented divisions at low imposed stress, as shown in Fig.~\ref{fig:EPM_division_avalanches}.

\subsubsection{Stress-oriented divisions}

\begin{figure*}
    \centering
    \includegraphics[width=0.99\linewidth]{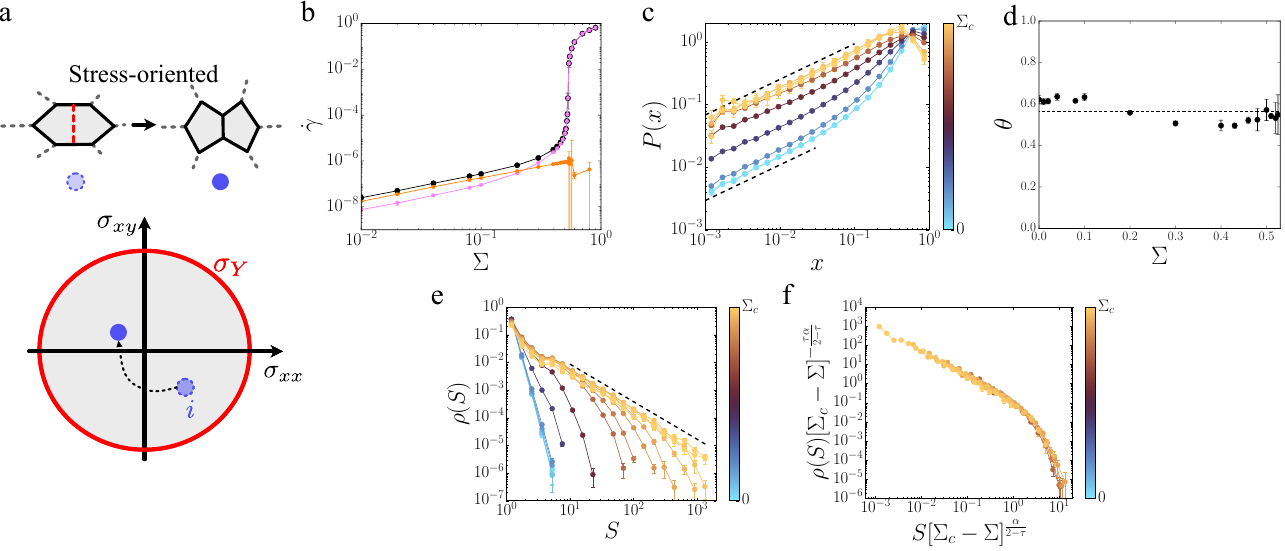}
    \caption{The main results of the main text studied for the oriented cell divisions. (a) Motivated by experimental data, the axis of division is chosen to align with the cell's local stress. (b) The decomposition of  the flow shows contributions from  both divisions $C$ and rearrangements $R$. (c) The distribution $P(x)$ exhibits a pseudogap. (d) The obtained value of the pseudogap exponent by fitting $P(x)\sim x^\theta$ for $0.001 < x < 0.03$. The dashed line shows an average over $\Sigma$ with $\theta=0.55 \pm 0.05$. (e, f) The avalanche size distribution follows a power law decay $\rho{S} \sim S^{-\tau}$, with $\tau=1.36$ shown by the black dashed line. Stress-dependent cutoff size is identified to follow $S_c \sim \left[\Sigma_c - \Sigma\right]^{\alpha/2-\tau}$, where similar to before, we find $\alpha=1.00\pm 0.05$. In panels (e, f), the system size is $L=256$, and in panel (b), the simulation size is $L=128$, and the relaxed stress is drawn from the disribution defined in Eq. A3.}
    \label{fig:oriented_divisions}
\end{figure*}

The stress-oriented divisions occur along the axis set by  the local stress, but the magnitude of the imprinted strain is random, with $\mathbf{E}[{\sigma}_m] = \Delta_{\text{div}}$ and $\mathbf{E}[{\sigma}^2_m] = \xi_{\text{div}}$. The average elastic energy change per division is given by
\begin{align}
    \mathbf{E}[\delta E_m^{\text{div}}] &= \left(\frac{1}{2}+D\right)\xi_{\text{div}} + {\Sigma} \, \Delta_{xy,m} \, \mathbf{E}\left[\frac{|{\sigma}_{xy,i_m}|}{{\sigma}_{xy,i_m}}\right] \nonumber\\
    &\quad - \mathbf{E}[|{\sigma}_{xx,i_m}|]\Delta_{\text{div},xx} - \mathbf{E}[|{\sigma}_{xy,i_m}|]\Delta_{\text{div},xy}. \label{eq:elastic_E_oriented_div}
\end{align}
Unlike the previous division types, the elastic energy change for stress-oriented divisions is more complex. 

As we showed earlier, the energy relaxation of rearrangements vanishes as $\Sigma_c - \Sigma$ in the limit of $\Sigma \to \Sigma_c^-$. In this limit, we expect that the properties of the flow will be determined by system-spanning avalanches of rearrangements. Thus, all division types are expected to behave similarly near $\Sigma_c$. 
However, far from $\Sigma_c$, the division mechanism can become an important factor. Then, to gain insight about the influence of stress-oriented divisions, we focus on the limit of ${\Sigma}=0$. In this limit,
\begin{align}
    \mathbf{E}[\delta E_m^{\text{div}}] = \left(\frac{1}{2}+D\right)\xi_{\text{div}} &- \mathbf{E}[|{\sigma}_{xx,i_m}|]\Delta_{\text{div},xx} \nonumber \\
    -&\mathbf{E}[|{\sigma}_{xy,i_m}|]\Delta_{\text{div},xy}. \label{eq:elastic_E_oriented_div_zero}
\end{align}
Stress-oriented divisions inject elastic energy if the right-hand side of Eq.~\ref{eq:elastic_E_oriented_div_zero} is positive, i.e.
\begin{align}
    \xi_\text{div} \geq \frac{\mathbf{E}[|{\sigma}_{xx,i_m}|]\Delta_{\text{div},xx} + \mathbf{E}[|{\sigma}_{xy,i_m}|]\Delta_{\text{div},xy}}{\frac{1}{2}+D}.
\end{align}

To further simplify the system, we consider an isotropic case where the stress components have similar expected values, i.e. $\mathbf{E}[|{\sigma}_{xx,i_m}|] = \mathbf{E}[|{\sigma}_{xy,i_m}|] = \mathbf{E}[|{\sigma}_{i_m}|]$. Using traingle inequality $\Delta_{\text{div},xx}+\Delta_{\text{div},xy} \geq \Delta_{\text{div}}$, we find,
\begin{align}
    \frac{\xi_{\text{div}}}{\Delta_{\text{div}}} \geq \frac{\mathbf{E}[|{\sigma}_{i_m}|]}{\frac{1}{2}+D}.
\end{align}
When this holds, stress-oriented divisions can inject energy even at low imposed stress and thereby sustain substantial mechanically triggered avalanches. Therefore, the stress-oriented divisions can both inject or relax energy at low shear stress, depending on the relative magnitude of the first and the second moments of the imprinted strain distributions of cell divisions $\xi_\text{div}/\Delta_{\text{div}}$. 

To illustrate this prediction, we incorporate stress-oriented divisions into EPM by choosing parameters such that the energy injection by stress-oriented divisions vanishes at low stress. In this regime, we expect that cell divisions will be the main contributor to the flow, yet cell rearrangements should not be removed totally. The simulation results confirm this as the obtained flow curve of stress-oriented divisions has contributions from both $C_{ij}$ and $R_{ij}$, as shown in Fig.~\ref{fig:oriented_divisions} (b). Moreover, the distribution of $P(x)$ does not show a depletion of low local stabilities as shown in Fig.~\ref{fig:oriented_divisions} (c) and shows a pseudogap exponent of $\theta=0.55\pm 0.05$, see Fig.~\ref{fig:oriented_divisions} (d). Finally, the avalanche size distribution follows a power law with a stress-dependent cutoff size, where, similar to before, we find $\alpha=1.00\pm 0.05$, as shown in Fig.~\ref{fig:oriented_divisions} (e, f). 

Overall, the elastic energy balance provides a compact description of the EPM energy dynamics, directly relating energy injection and relaxation by cellular processes to the statistics of plastic avalanches. In particular, it yields a quantitative prediction for the mean avalanche size, consistent with our numerical measurements.

\end{document}